\documentclass[11pt]{article}
\usepackage{graphicx}
\usepackage{amsmath}
\usepackage{amssymb}
\usepackage{caption2}
\begin{document}
\bibliographystyle{prsty}
\begin{center}
{\large {\bf \sc{  Analysis of  the ${1\over 2}^+$ doubly heavy
baryon states   with QCD sum rules }}} \\[2mm]
Zhi-Gang Wang \footnote{E-mail,wangzgyiti@yahoo.com.cn.  }     \\
 Department of Physics, North China Electric Power University,
Baoding 071003, P. R. China

\end{center}

\begin{abstract}
In this article, we study the  ${1\over 2}^+$ doubly heavy baryon
states $\Omega_{QQ}$ and $\Xi_{QQ}$  by subtracting the
contributions from the corresponding ${1\over 2}^-$ doubly heavy
baryon states with the QCD sum rules, and make reasonable
predictions for their masses.  Those doubly heavy baryon states
maybe observed at the Tevatron, the LHCb and the PANDA.
\end{abstract}



\section{Introduction}

In 2002,  the SELEX collaboration reported   the first observation
of a signal for the doubly charmed baryon state  $ \Xi_{cc}^+$ in
the charged decay mode $\Xi_{cc}^+\rightarrow\Lambda_c^+K^-\pi^+$
\cite{SELEX2002}, and confirmed later by the same collaboration in
the decay mode $\Xi_{cc}^+\rightarrow pD^+K^- $ with the measured
mass $M_{\Xi}=(3518.9 \pm 0.9) \,\rm{ MeV }$ \cite{SELEX2004}.
However, the Babar and Belle collaborations  have not observed any
evidence for the doubly charmed baryons in $e^{+}e^{-}$
annihilations \cite{Xi-cc-1,Xi-cc-2}.  The charmed  and bottom
baryons which contain one (two) heavy quark(s) are particularly
interesting for studying dynamics of the light quarks in the
presence  of the heavy quark(s), and serve as an excellent ground
for testing predictions of the
 quark models and heavy quark symmetry. There have been several
approaches to deal with  the doubly heavy baryon masses, such as the
relativistic quark model \cite{HH-Ebert,HH-Martynenko}, the
non-relativistic quark model
\cite{HH-Roberts,HH-Albertus,HH-Vijande,HH-Gershtein}, the
three-body Faddeev method \cite{HH-Valcarce}, the potential approach
combined with the QCD sum rules \cite{HH-Kiselev}, the quark model
with AdS/QCD inspired potential \cite{HH-Giannuzzi}, the MIT bag
model \cite{HH-He}, the full QCD sum rules
\cite{HH-Narison,HH-Zhang}, and the Feynman-Hellmann theorem and
semiempirical mass formulas \cite{Massform}, etc.

The QCD sum rules is a powerful theoretical tool in studying the
ground state heavy baryons \cite{SVZ79,PRT85,NarisonBook}. In the
QCD sum rules, the operator product expansion is used to expand the
time-ordered currents into a series of quark and gluon condensates
which parameterize the long distance properties of the QCD vacuum.
Based on the quark-hadron duality, we can obtain copious information
about the hadronic parameters at the phenomenological side
\cite{SVZ79,PRT85}. There have been several works on the masses of
the heavy baryon states with the full QCD sum rules and the QCD sum
rules in the heavy quark effective theory (one can consult
Ref.\cite{Wang0912} for more literatures).

In Ref.\cite{Oka96}, Jido et al introduce a novel approach based on
the QCD sum rules to separate the contributions of   the negative
parity light flavor  baryon states from the positive parity light
flavor baryon states, as the interpolating currents may have
non-vanishing couplings to both the negative parity and positive
parity baryon states \cite{Chung82}. Before the work of Jido et al,
Bagan et al take the infinite mass limit for the heavy quarks to
separate the contributions of the positive parity and negative
parity heavy baryon states unambiguously \cite{Bagan93}.

 In
Refs.\cite{Wang0704,Wang0809,Wang0910}, we study the
 heavy baryon states $\Omega_Q$, $\Xi'_Q$, $\Sigma_Q$, $\Omega_Q^*$, $\Xi^*_Q$ and
$\Sigma^*_Q$ with the full QCD sum rules, and observe that the pole
residues of the ${\frac{3}{2}}^+$ heavy baryon states from the sum
rules with different tensor structures are consistent with each
other, while the pole residues of the ${\frac{1}{2}}^+$ heavy baryon
states from the sum rules with different tensor structures differ
from each other greatly.  In Refs.\cite{Wang0912,Wang0101}, we
follow Ref.\cite{Oka96} and study the masses and pole residues of
the ${\frac{1}{2}}^+$ heavy baryon states $\Omega_Q$, $\Xi'_Q$,
$\Sigma_Q$, $\Lambda_Q$ and $\Xi_Q$ by subtracting the contributions
of  the negative parity heavy baryon states to overcome  the
embarrassment. Those pole residues are important parameters in
studying the radiative decays $\Omega_Q^*\to \Omega_Q \gamma$,
$\Xi_Q^*\to \Xi'_Q \gamma$ and $\Sigma_Q^*\to \Sigma_Q \gamma$
\cite{Wang0910,Wang0909}, etc.

In this article, we extend our previous works  to study the ${1\over
2}^+$ doubly heavy baryon states $\Xi_{QQ}$ and $\Omega_{QQ}$ with
the QCD sum rules.

The article is arranged as follows:  we derive the QCD sum rules for
the masses and the pole residues of  the doubly heavy baryon states
$\Xi_{QQ}$ and  $\Omega_{QQ}$  in Sec.2; in Sec.3, we present the
 numerical results and discussions; and Sec.4 is reserved for our
conclusions.

\section{QCD sum rules for  the $\Xi_{QQ}$ and  $\Omega_{QQ}$ }
The ${1\over 2}^+$ doubly heavy baryon states $\Xi_{QQ}$ and
$\Omega_{QQ}$ can be interpolated by the following currents
$J_\Xi(x)$ and $J_\Omega(x)$
 respectively,
\begin{eqnarray}
J_\Xi(x)&=& \epsilon^{ijk}  Q^T_i(x)C\gamma_\mu Q_j(x)
\gamma_5\gamma^\mu q_k(x)  \, ,  \nonumber \\
 J_\Omega(x)&=& \epsilon^{ijk}  Q^T_i(x)C\gamma_\mu Q_j(x)   \gamma_5\gamma^\mu s_k(x) \, ,
\end{eqnarray}
where the  $Q$ represents the heavy quarks $c$ and $b$,  the $i$,
$j$ and $k$ are color indexes, and the $C$ is the charge conjunction
matrix. In the heavy quark limit, the doubly heavy baryon states can
be described by the  (heavy)diquark-(light)quark model
\cite{HH-Kiselev}.

 The corresponding ${1\over 2}^-$ doubly heavy baryon states can be
interpolated by the  currents $J_{-} =i\gamma_{5} J_{+}$  because
multiplying $i \gamma_{5}$ to $J_{+}$ changes the parity of $J_{+}$
\cite{Oka96}, where the $J_{+}$ denotes the currents $J_\Omega(x)$
and $J_\Xi(x)$. The correlation functions are defined by
\begin{eqnarray}
\Pi_{\pm}(p)&=&i\int d^4x e^{ip \cdot x} \langle
0|T\left\{J_{\pm}(x)\bar{J}_{\pm}(0)\right\}|0\rangle \, ,
\end{eqnarray}
and can be decomposed as
\begin{equation}
    \Pi_{\pm}(p) = \!\not\!{p} \Pi_{1}(p^{2}) \pm \Pi_{0}(p^{2})\, ,
\end{equation}
due to Lorentz covariance. The currents $J_{+}$ couple  to both the
positive parity  and negative parity baryon states \cite{Chung82},
i.e. $ \langle{0}|J_{+}| B^{-}\rangle \langle
B^{-}|\bar{J}_{+}|0\rangle =
    - \gamma_{5}\langle 0|J_{-}| B^{-}\rangle \langle B^{-}| \bar{J}_{-}|0\rangle
    \gamma_{5}$, where the $B^{-}$ denote the negative parity baryon states.

 We  insert  a complete set  of intermediate baryon states with the same quantum
numbers as the current operators $J_{+}(x)$ and $J_{-}(x)$ into the
correlation functions $\Pi_{+}(p)$  to obtain the hadronic
representation \cite{SVZ79,PRT85}. After isolating the pole terms of
the lowest states of the doubly heavy baryons, we obtain the
following result \cite{Oka96}:
\begin{eqnarray}
    \Pi_{+}(p)     & = &   \lambda_+^2 {\!\not\!{p} +
    M_{+} \over M^{2}_+ -p^{2} } + \lambda_{-}^2
    {\!\not\!{p} - M_{-} \over M_{-}^{2}-p^{2}  } +\cdots \, ,
    \end{eqnarray}
where the $M_{\pm}$ are the masses of the lowest states with parity
$\pm$ respectively, and the $\lambda_{\pm}$ are the  corresponding
pole residues (or couplings).
 If we take $\vec{p} = 0$, then
\begin{eqnarray}
  \rm{limit}_{\epsilon\rightarrow0}\frac{{\rm Im}  \Pi_+(p_{0}+i\epsilon)}{\pi} & = &
    \lambda_+^2 {\gamma_{0} + 1\over 2} \delta(p_{0} - M_+) +
    \lambda_{-}^{2} {\gamma_{0} - 1\over 2} \delta(p_{0} - M_{-})+\cdots \nonumber \\
  & = & \gamma_{0} A(p_{0}) + B(p_{0})+\cdots \, ,
\end{eqnarray}
where
\begin{eqnarray}
  A(p_{0}) & = & {1 \over 2} \left[ \lambda_+^{2}
  \delta(p_{0} - M_+)  + \lambda_-^{2} \delta(p_{0} -
  M_{-})\right] \, , \nonumber \\
   B(p_{0}) & = & {1 \over 2} \left[ \lambda_+^{2}
  \delta(p_{0} - M_+)  - \lambda_-^{2} \delta(p_{0} -
  M_{-})\right] \, ,
\end{eqnarray}
the  $A(p_{0}) + B(p_{0})$ and $A(p_{0}) - B(p_{0})$ contain the
contributions  from the positive-parity and negative-parity baryon
states respectively.

We  calculate the light quark parts of the correlation functions
$\Pi_{+}(p)$ in the coordinate space and use the momentum space
expression for the heavy quark propagators, i.e. we take
\begin{eqnarray}
S_{ij}(x)&=& \frac{i\delta_{ij}\!\not\!{x}}{ 2\pi^2x^4}
-\frac{\delta_{ij}m_s}{4\pi^2x^2}-\frac{\delta_{ij}}{12}\langle
\bar{s}s\rangle +\frac{i\delta_{ij}}{48}m_s
\langle\bar{s}s\rangle\!\not\!{x}     \nonumber\\
&& -\frac{i}{32\pi^2x^2}  G^{ij}_{\mu\nu}(x) \left[\!\not\!{x}
\sigma^{\mu\nu}+\sigma^{\mu\nu} \!\not\!{x}\right]  +\cdots \, ,\nonumber\\
S_Q^{ij}(x)&=&\frac{i}{(2\pi)^4}\int d^4k e^{-ik \cdot x} \left\{
\frac{\delta_{ij}}{\!\not\!{k}-m_Q}
-\frac{g_sG^{\alpha\beta}_{ij}}{4}\frac{\sigma_{\alpha\beta}(\!\not\!{k}+m_Q)+(\!\not\!{k}+m_Q)
\sigma_{\alpha\beta}}{(k^2-m_Q^2)^2}\right.\nonumber\\
&&\left.+\frac{\pi^2}{3} \langle \frac{\alpha_sGG}{\pi}\rangle
\delta_{ij}m_Q \frac{k^2+m_Q\!\not\!{k}}{(k^2-m_Q^2)^4}
+\cdots\right\} \, ,
\end{eqnarray}
where $\langle \frac{\alpha_sGG}{\pi}\rangle=\langle
\frac{\alpha_sG_{\alpha\beta}G^{\alpha\beta}}{\pi}\rangle$, then
resort to the Fourier integral to transform  the light quark parts
into the momentum space in $D$ dimensions,  take $\vec{p} = 0$,  and
use the dispersion relation to obtain the spectral densities
$\rho^A(p_0)$ and $\rho^B(p_0)$ (which correspond to the tensor
structures $\gamma_0$ and $1$ respectively) at the level of
quark-gluon degrees of freedom. Finally we introduce the weight
functions $\exp\left[-\frac{p_0^2}{T^2}\right]$,
$p_0^2\exp\left[-\frac{p_0^2}{T^2}\right]$,   and obtain the
following sum rules,
\begin{eqnarray}
  \lambda_{+}^2\exp\left[-\frac{M_+^2}{T^2}\right]&=&\int_{\Delta}^{\sqrt{s_0}}dp_0
\left[\rho^A(p_0)
+\rho^B(p_0)\right]\exp\left[-\frac{p_0^2}{T^2}\right] \, ,
\end{eqnarray}
\begin{eqnarray}
  \lambda_{+}^2M_+^2\exp\left[-\frac{M_+^2}{T^2}\right]&=&\int_{\Delta}^{\sqrt{s_0}}dp_0
\left[\rho^A(p_0)
+\rho^B(p_0)\right]p_0^2\exp\left[-\frac{p_0^2}{T^2}\right] \, ,
\end{eqnarray}
where
\begin{eqnarray}
\rho^A_{\Omega}(p_0)&=&\frac{3p_0}{8 \pi^4}
\int_{\alpha_{i}}^{\alpha_{f}}d\alpha \int_{\beta_{i}}^{1-\alpha}
d\beta\alpha\beta(1-\alpha-\beta)(p_0^2-\widetilde{m}^2_Q)(5p_0^2-3\widetilde{m}^2_Q)
\nonumber\\
&&+\frac{3m_Q^2p_0}{8\pi^4}\int_{\alpha_{i}}^{\alpha_{f}}d\alpha
\int_{\beta_{i}}^{1-\alpha} d\beta
(1-\alpha-\beta)(p_0^2-\widetilde{m}^2_Q) \nonumber\\
&&-\frac{m_Q^2}{24\pi^2}
\langle\frac{\alpha_sGG}{\pi}\rangle\int_{\alpha_{i}}^{\alpha_{f}}d\alpha
\int_{\beta_{i}}^{1-\alpha} d\beta (1-\alpha-\beta)
\left[\frac{\alpha}{\beta^2}+\frac{\beta}{\alpha^2} \right]\left[1+\frac{p_0}{4T}\right]\delta(p_0-\widetilde{m}_Q)\nonumber\\
&&-\frac{m_Q^4}{192\pi^2p_0T}\langle\frac{\alpha_sGG}{\pi}\rangle\int_{\alpha_{i}}^{\alpha_{f}}d\alpha
\int_{\beta_{i}}^{1-\alpha} d\beta (1-\alpha-\beta)\left[\frac{1}{\alpha^3}+\frac{1}{\beta^3}\right]\delta(p_0-\widetilde{m}_Q)\nonumber\\
&&+\frac{m_Q^2}{32\pi^2}\langle\frac{\alpha_sGG}{\pi}\rangle\int_{\alpha_{i}}^{\alpha_{f}}d\alpha
\int_{\beta_{i}}^{1-\alpha} d\beta (1-\alpha-\beta)\left[\frac{1}{\alpha^2}+\frac{1}{\beta^2}\right]\delta(p_0-\widetilde{m}_Q)\nonumber\\
&&+\frac{m_s\langle\bar{s}{s}\rangle}{4\pi^2}\int_{\alpha_{i}}^{\alpha_{f}}d\alpha
\alpha(1-\alpha)\left[
6p_0+p_0^2\delta(p_0-\widetilde{\widetilde{m}}_Q)\right] \nonumber\\
&&+\frac{m_sm_Q^2\langle\bar{s}{s}\rangle}{8\pi^2}\int_{\alpha_{i}}^{\alpha_{f}}d\alpha
\delta(p_0-\widetilde{\widetilde{m}}_Q)  \nonumber \\
&&+\frac{1}{32\pi^2}\langle\frac{\alpha_sGG}{\pi}\rangle\int_{\alpha_{i}}^{\alpha_{f}}d\alpha
\int_{\beta_{i}}^{1-\alpha} d\beta (\alpha+\beta)
\left[3p_0+\frac{\widetilde{m}_Q^2}{2}\delta(p_0-\widetilde{m}_Q)
\right]\nonumber \\
&&+\frac{m_Q^2}{64\pi^2}\langle\frac{\alpha_sGG}{\pi}\rangle\int_{\alpha_{i}}^{\alpha_{f}}d\alpha
\int_{\beta_{i}}^{1-\alpha} d\beta \frac{\alpha+\beta}{\alpha\beta}
 \delta(p_0-\widetilde{m}_Q)\, ,
\end{eqnarray}
\begin{eqnarray}
\rho^B_{\Omega}(p_0)&=&\frac{3m_s}{8 \pi^4}
\int_{\alpha_{i}}^{\alpha_{f}}d\alpha \int_{\beta_{i}}^{1-\alpha}
d\beta\alpha\beta(p_0^2-\widetilde{m}^2_Q)(2p_0^2-\widetilde{m}^2_Q)
\nonumber\\
&&+\frac{3m_sm_Q^2}{4\pi^4}\int_{\alpha_{i}}^{\alpha_{f}}d\alpha
\int_{\beta_{i}}^{1-\alpha} d\beta
(p_0^2-\widetilde{m}^2_Q) \nonumber\\
&&-\frac{m_sm_Q^2}{96\pi^2}
\langle\frac{\alpha_sGG}{\pi}\rangle\int_{\alpha_{i}}^{\alpha_{f}}d\alpha
\int_{\beta_{i}}^{1-\alpha} d\beta
\left[\frac{\alpha}{\beta^2}+\frac{\beta}{\alpha^2} \right]\left[\frac{1}{\widetilde{m}_Q}+\frac{1}{2T}\right]\delta(p_0-\widetilde{m}_Q)\nonumber\\
&&-\frac{m_sm_Q^4}{96\pi^2p_0^2T}\langle\frac{\alpha_sGG}{\pi}\rangle\int_{\alpha_{i}}^{\alpha_{f}}d\alpha
\int_{\beta_{i}}^{1-\alpha} d\beta \left[\frac{1}{\alpha^3}+\frac{1}{\beta^3}\right]\delta(p_0-\widetilde{m}_Q)\nonumber\\
&&+\frac{m_sm_Q^2}{16\pi^2p_0}\langle\frac{\alpha_sGG}{\pi}\rangle\int_{\alpha_{i}}^{\alpha_{f}}d\alpha
\int_{\beta_{i}}^{1-\alpha} d\beta \left[\frac{1}{\alpha^2}+\frac{1}{\beta^2}\right]\delta(p_0-\widetilde{m}_Q)\nonumber\\
&&-\frac{\langle\bar{s}{s}\rangle}{2\pi^2}\int_{\alpha_{i}}^{\alpha_{f}}d\alpha
\alpha(1-\alpha)\left[ 3p_0^2-2\widetilde{\widetilde{m}}_Q^2\right]
-\frac{m_Q^2\langle\bar{s}{s}\rangle}{\pi^2}\int_{\alpha_{i}}^{\alpha_{f}}d\alpha
 \nonumber \\
&&-\frac{m_s}{16\pi^2}\langle\frac{\alpha_sGG}{\pi}\rangle\int_{\alpha_{i}}^{\alpha_{f}}d\alpha
\int_{\beta_{i}}^{1-\alpha} d\beta
\left[1+\frac{\widetilde{m}_Q}{4}\delta(p_0-\widetilde{m}_Q) \right]
\, ,
\end{eqnarray}
the $s_0$ are the threshold parameters, $T^2$ are the Borel
parameters, $\alpha_{f}=\frac{1+\sqrt{1-4m_Q^2/p_0^2}}{2}$,
$\alpha_{i}=\frac{1-\sqrt{1-4m_Q^2/p_0^2}}{2}$,
$\beta_{i}=\frac{\alpha m_Q^2}{\alpha p_0^2 -m_Q^2}$,
$\widetilde{m}_Q^2=\frac{(\alpha+\beta)m_Q^2}{\alpha\beta}$,
$\widetilde{\widetilde{m}}_Q^2=\frac{m_Q^2}{\alpha(1-\alpha)}$,
 and $\Delta=2m_Q+m_s$.
With the simple replacements, $\langle\bar{s}s\rangle
\rightarrow\langle\bar{q}q\rangle$ and $m_s\rightarrow 0$,  we can
obtain the corresponding spectral densities  of the $\Xi_{QQ}$ at
the level of quark-gluon degrees of freedom.

\section{Numerical results and discussions}
The input parameters are taken as $\langle \bar{q}q
\rangle=-(0.24\pm 0.01 \,\rm{GeV})^3$,  $\langle \bar{s}s
\rangle=(0.8\pm 0.2 )\langle \bar{q}q \rangle$, $\langle
\bar{q}g_s\sigma Gq \rangle=m_0^2\langle \bar{q}q \rangle$, $\langle
\bar{s}g_s\sigma Gs \rangle=m_0^2\langle \bar{s}s \rangle$,
$m_0^2=(0.8 \pm 0.2)\,\rm{GeV}^2$ \cite{Ioffe2005,LCSRreview},
$\langle \frac{\alpha_s GG}{\pi}\rangle=(0.012 \pm
0.004)\,\rm{GeV}^4 $ \cite{LCSRreview},
$m_s=(0.14\pm0.01)\,\rm{GeV}$, $m_c=(1.35\pm0.10)\,\rm{GeV}$ and
$m_b=(4.7\pm0.1)\,\rm{GeV}$ \cite{PDG} at the energy scale  $\mu=1\,
\rm{GeV}$.

The value of the gluon condensate $\langle \frac{\alpha_s
GG}{\pi}\rangle $ has been updated from time to time, and changes
greatly \cite{NarisonBook}.
 At the present case, the gluon condensate  makes tiny  contribution (see Table 1),
the updated value $\langle \frac{\alpha_s GG}{\pi}\rangle=(0.023 \pm
0.003)\,\rm{GeV}^4 $ \cite{NarisonBook} and the standard value
$\langle \frac{\alpha_s GG}{\pi}\rangle=(0.012 \pm
0.004)\,\rm{GeV}^4 $ \cite{LCSRreview} lead to a slight difference
and can be neglected safely.

The $Q$-quark masses appearing in the perturbative terms  are
usually taken to be the pole masses in the QCD sum rules, while the
choice of the $m_Q$ in the leading-order coefficients of the
higher-dimensional terms is arbitrary \cite{NarisonBook,Kho9801}.
The $\overline{MS}$ mass $m_c(m_c^2)$ relates with the pole mass
$\hat{m}_c$ through the relation $ m_c(m_c^2)
=\hat{m}_c\left[1+\frac{C_F \alpha_s(m_c^2)}{\pi}+\cdots\right]^{-1}
$. In this article, we take the approximation
$m_c(m_c^2)\approx\hat{m}_c$ without the $\alpha_s$ corrections for
consistency. The value listed in the Particle Data Group is
$m_c(m_c^2)=1.27^{+0.07}_{-0.11} \, \rm{GeV}$ \cite{PDG}, it is
reasonable to take
$\hat{m}_c=m_c(1\,\rm{GeV}^2)=(1.35\pm0.10)\,\rm{GeV}$. For the $b$
quark,  the $\overline{MS}$ mass
$m_b(m_b^2)=4.20^{+0.17}_{-0.07}\,\rm{GeV}$ \cite{PDG}, the
  gap between the energy scale $\mu=4.2\,\rm{GeV}$ and
 $1\,\rm{GeV}$ is rather large, the approximation $\hat{m}_b\approx m_b(m_b^2)\approx m_b(1\,\rm{GeV}^2)$ seems rather crude.
  It would be better to understand the quark masses $m_c$ and $m_b$ we
take at the energy scale $\mu^2=1\,\rm{GeV}^2$ as the effective
quark masses (or just the mass parameters).

 In calculation, we   neglect  the contributions from the
perturbative $\mathcal {O}(\alpha_s^n)$  corrections.  Those
perturbative corrections can be taken into account in the leading
logarithmic  approximations through the anomalous dimension factors.
After the Borel transform, the effects of those
 corrections are  to multiply each term on the operator product
 expansion side by the factor, $ \left[ \frac{\alpha_s(T^2)}{\alpha_s(\mu^2)}\right]^{2\Gamma_{J}-\Gamma_{\mathcal
 {O}_n}}  $,
 where the $\Gamma_{J}$ is the anomalous dimension of the
 interpolating current $J(x)$, and the $\Gamma_{\mathcal {O}_n}$ is the anomalous dimension of
 the local operator $\mathcal {O}_n(0)$, which
governs the evolution of the vacuum condensate
$\langle{O}_n(0)\rangle_\mu$ with the energy scale through the
renormalization group equation.

 If the perturbative
$\mathcal {O}(\alpha_s)$ corrections and the anomalous dimension
factors are taken into account consistently, the spectral densities
in the QCD side should be replaced with
\begin{eqnarray}
\mathcal {O}_0(0) &\rightarrow &\left[
\frac{\alpha_s(T^2)}{\alpha_s(\mu^2)}\right]^{2\Gamma_{J}}
\left[1+A(p_0^2,m_Q^2)\frac{\alpha_s(T^2)}{\pi} \right]\mathcal
{O}_0(0) \, ,\nonumber \\
\langle\mathcal {O}_n(0)\rangle_{\mu} &\rightarrow &\left[
\frac{\alpha_s(T^2)}{\alpha_s(\mu^2)}\right]^{2\Gamma_{J}-\Gamma_{\mathcal
 {O}_n} }\left[1+B(p_0^2,m_Q^2)\frac{\alpha_s(T^2)}{\pi} \right]\langle\mathcal{O}_n(0)\rangle_{\mu} \, ,\nonumber
\end{eqnarray}
where the $A(p_0^2,m_Q^2)$ and $B(p_0^2,m_Q^2)$ are some notations
for the coefficients of the perturbative corrections, the average
virtuality of the quarks in the correlation functions is
 characterized by the Borel parameter $T^2$. We cannot estimate the
corrections and the uncertainties originate from the corrections
with confidence without explicit calculations.   In the case of the
correlation function for  the proton,  the perturbative $\mathcal
{O}(\alpha_s)$ corrections can change
 the numerical values of the mass and the pole residue
 considerably and improve the agreement between the
theoretical estimation and the experimental data
\cite{Ioffe2005,Sulian}. In the present case, including the
$\alpha_s$ corrections maybe improve the predictions.

In this article, we carry out the operator product expansion at the
special energy scale $\mu^2=1\,\rm{GeV}^2$, and  set the factor
$\left[\frac{\alpha_s(T^2)}{\alpha_s(\mu^2)}\right]^{2\Gamma_{J}-\Gamma_{\mathcal
{O}_n}}\approx1$ for consistency, as the $\alpha_s$ corrections have
not been calculated yet.  Such an approximation maybe result in some
scale dependence and weaken the prediction ability. In this article,
we study the $J^P=\frac{1}{2}^+$ doubly heavy baryon states
systemically,  the predictions are   reasonable as we take the
analogous criteria in those sum rules.

In the conventional QCD sum rules \cite{SVZ79,PRT85}, there are two
criteria (pole dominance and convergence of the operator product
expansion) for choosing  the Borel parameter $T^2$ and threshold
parameter $s_0$.  We impose the two criteria on the doubly heavy
baryon states to choose the Borel parameter $T^2$ and threshold
parameter $s_0$.

\begin{figure}
 \centering
 \includegraphics[totalheight=5cm,width=6cm]{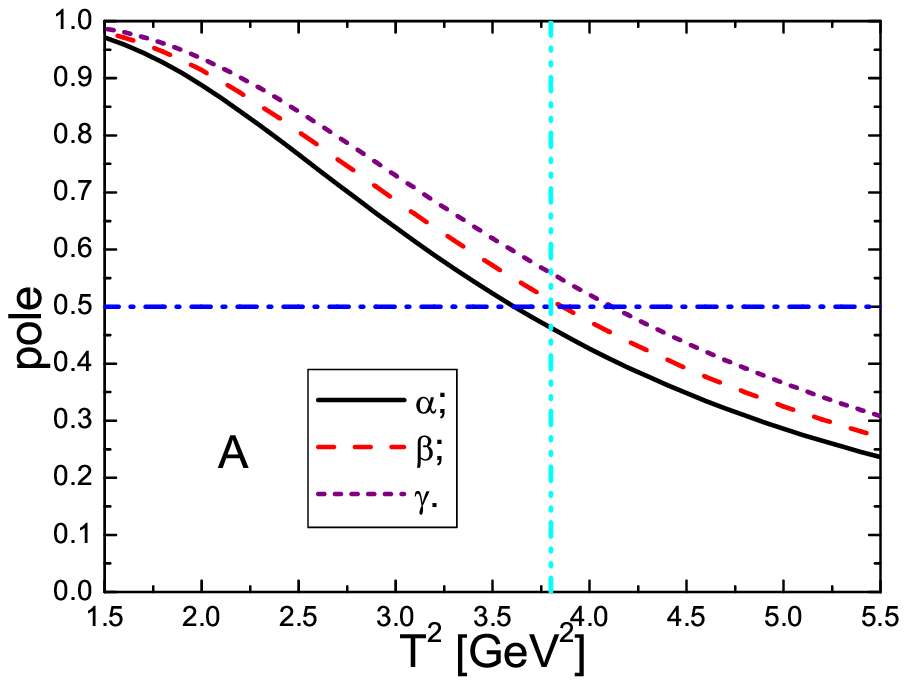}
 \includegraphics[totalheight=5cm,width=6cm]{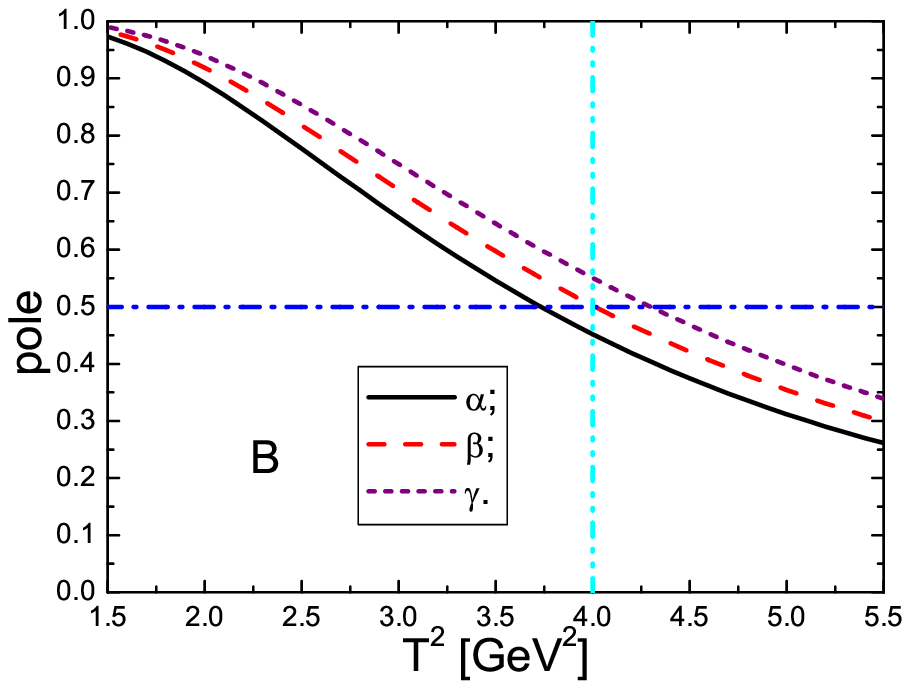}
\includegraphics[totalheight=5cm,width=6cm]{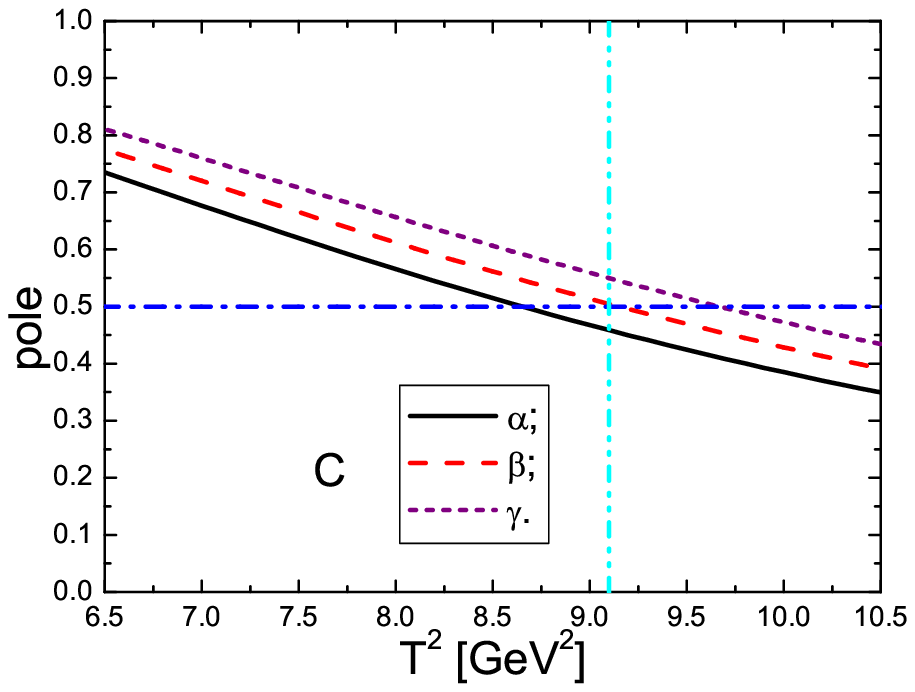}
\includegraphics[totalheight=5cm,width=6cm]{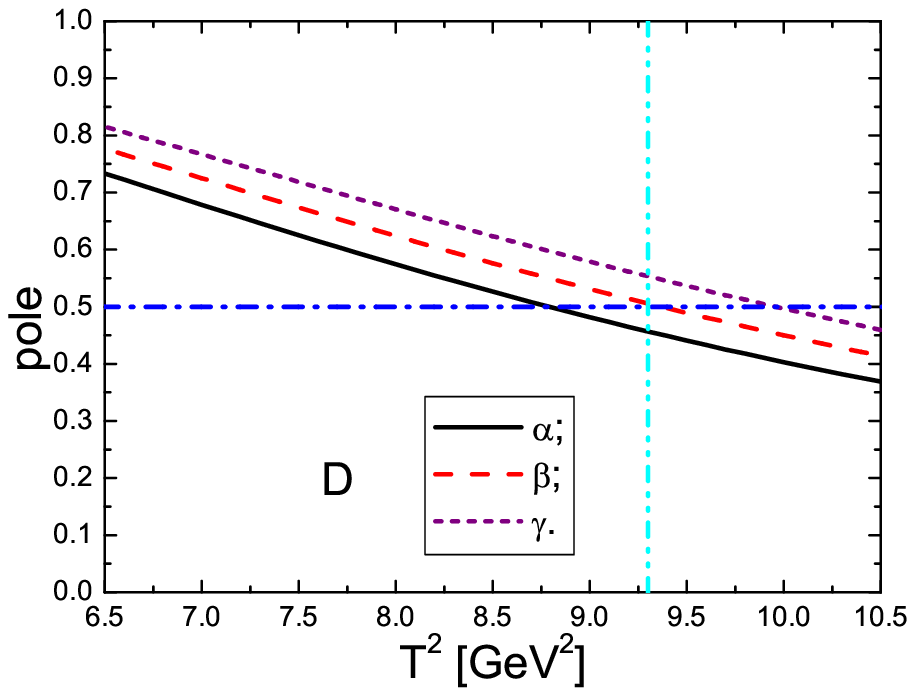}
        \caption{ The contributions of the pole terms with variations of the Borel parameters $T^2$, the $A$, $B$, $C$ and $D$  correspond
      to the channels $\Xi_{cc}$, $\Omega_{cc}$, $\Xi_{bb}$ and $\Omega_{bb}$ respectively, the $\beta$ corresponds to the
central values of the threshold parameters, the energy gap among
$\alpha$, $\beta$ and $\gamma$ is $0.1\,\rm{GeV}$. }
\end{figure}

In Fig.1, we plot the  contributions from the pole terms with
variations  of the Borel parameters $T^2$ and the threshold
parameters $s_0$. The pole contributions are larger than (or about)
$50\%$ at the values which are denoted by the vertical lines for
central values ($\beta$) of the threshold parameters $s_0$. From the
figure, we can set the upper bound of the  Borel parameters
$T^2_{max}$,  $T^2_{max}=3.8\,\rm{GeV}^2$, $4.0\,\rm{GeV}^2$,
$9.1\,\rm{GeV}^2$ and $9.3\,\rm{GeV}^2$ in the channels $\Xi_{cc}$,
$\Omega_{cc}$, $\Xi_{bb}$ and $\Omega_{bb}$, respectively.
  The lower bound of the Borel parameters $T^2_{min}$
can be determined  in the regions  where the contributions from the
perturbative terms are larger than (or equal) the ones from the
quark condensates, $T^2_{min}= 2.8\,\rm{GeV}^2$, $1.6\,\rm{GeV}^2$,
$7.7\,\rm{GeV}^2$ and $<6.5\,\rm{GeV}^2$ in the channels $\Xi_{cc}$,
$\Omega_{cc}$, $\Xi_{bb}$ and $\Omega_{bb}$, respectively. The
convergent behaviors in the channels $\Omega_{QQ}$ are better than
the corresponding ones in the channels $\Xi_{QQ}$, this is mainly
due to the fact that the values of the quark condensates, $|\langle
\bar{q}{q}\rangle|>|\langle \bar{s}{s}\rangle|$. The Borel windows
$T^2_{min}-T^2_{max}$ in the channels $\Xi_{QQ}$ and $\Omega_{QQ}$
overlap with each other. In this article, we can take the uniform
intervals for the Borel windows (IBW), i.e. $\rm{IBW}=1.0 \,
\rm{GeV}^2$ and $1.4 \, \rm{GeV}^2$ in the doubly charmed  and
doubly bottom channels respectively, which warrant the pole
contributions are analogous. The values of the threshold parameters
$s_0$ and the Borel parameters $T^2$ are shown in Table 1, from the
table, we can see that the two criteria of the QCD sum rules are
fully satisfied \cite{SVZ79,PRT85}.

In Ref.\cite{Wang0912}, we study the ${\frac{1}{2}}^+$ sextet  heavy
baryon states $\Omega_b$, $\Omega_c$, $\Xi'_b$, $\Xi'_c$, $\Sigma_b$
and $\Sigma_c$ by subtracting the contributions from the
corresponding negative parity heavy baryon states with the QCD sum
rules, the predicted masses are in good agreement with the
experimental data for the well established  mesons, $\Omega_b$,
$\Sigma_b$, $\Omega_c$, $\Xi'_c$ and $\Sigma_c$. In those sum rules,
the contributions from the pole terms are about $(45-65)\%$ and
$(45-80)\%$ for the bottom and  charmed baryon states respectively.
In this article,  we take analogous pole contributions, see Table 1,
the predictions are  reasonable although the contributions from the
perturbative continuum are somewhat large.

In Fig.2, we plot the  predicted masses with variations   of the
threshold parameters $s_0$. From the figure, we can see that the
predicted masses are not sensitive to the threshold parameters,
although they increase with the threshold parameters. In
calculation, we take uniform  uncertainties for the threshold
parameters, $\delta_{\sqrt{s_0}}=\pm 0.1\, \rm{GeV}$.

From Table 1, we can see that the contributions from the quark
condensates are large, even comparable with the perturbative terms,
this is an indication of the non-perturbative origin of the masses
of the baryon states. In the chiral limit, the spectral densities on
the QCD side come from the quark condensates only.  An astonishingly
simple expression can be obtained in case of the proton
\cite{Ioffe-1,Ioffe-2},
\begin{eqnarray}
M_p&=&\sqrt[3]{-8\pi^2\langle
\bar{q}q\rangle_{\mu=1\,\rm{GeV}}}\approx 1 \, \rm{GeV} \, .
\end{eqnarray}

Taking into account all uncertainties  of the relevant  parameters,
we can obtain the values of the masses and pole residues of
 the doubly  heavy baryon states
$\Xi_{QQ}$ and $\Omega_{QQ}$, which are shown in Figs.3-4 and Tables
2-3. In Table 2, we also present the predictions of other
theoretical approaches and the values of the experimental data, the
present predictions are consistent with them.

\begin{table}
\begin{center}
\begin{tabular}{|c|c|c|c|c|c|c|c|}
\hline\hline & $T^2 (\rm{GeV}^2)$& $\sqrt{s_0} (\rm{GeV})$&pole&perturbative& $\langle \bar{q}q\rangle$ & $\langle \frac{\alpha_sGG}{\pi}\rangle$\\
\hline
 $\Xi_{cc}$  &$2.8-3.8$ &$4.2$&  $(46-78)\%$&$(48-55)\%$ & $(43-50)\%$ &$\approx2\%$\\ \hline
   $\Omega_{cc}$  &$3.0-4.0$ &$4.3$& $(45-75)\%$&$(67-72)\%$& $(26-30)\%$ &$\approx2\%$\\ \hline
    $\Xi_{bb}$  &$7.7-9.1$ &$10.8$& $(46-69)\%$ &$(48-53)\%$& $(46-51)\%$&$<1\%$\\ \hline
    $\Omega_{bb}$  &$7.9-9.3$ &$10.9$ & $(46-68)\%$&$(69-72)\%$&$(28-31)\%$&$<1\%$\\ \hline \hline
\end{tabular}
\end{center}
\caption{ The Borel parameters $T^2$ and threshold parameters $s_0$
for the doubly heavy baryon states, the "pole" stands for the
contribution from the pole term, and the "perturbative" stands for
the contribution from the perturbative term in the operator product
expansion, etc. In calculating the contributions from the pole
terms, we take into account the uniform uncertainties of  the
threshold parameters, $\delta_{\sqrt{s_0}}=\pm 0.1\, \rm{GeV}$. }
\end{table}

\begin{table}
\begin{center}
\begin{tabular}{|c|c|c|c|c|}
\hline\hline References 
&$\Xi_{cc}$&$\Omega_{cc}$&$\Xi_{bb}$&$\Omega_{bb}$\\ \hline
\cite{HH-Ebert}& $3.620$&$3.778$&$10.202$ & $10.359$\\ \hline
\cite{HH-Roberts}& $3.676$ &$3.815$ &$10.340$ & $10.454$ \\ \hline
 \cite{HH-Albertus}& $3.612$ &$3.702$ &$10.197$ & $10.260$\\ \hline
  \cite{HH-Valcarce}&  $3.579$  &$3.697$ &$10.189$ &$10.293$\\ \hline
\cite{HH-Kiselev}& $3.48$ &$3.59$ &$10.09$ & $10.18$\\ \hline
 \cite{HH-Giannuzzi}&   $3.547$  &$3.648$ &$10.185$ & $10.271$\\ \hline
\cite{HH-He}& $3.520$ &$3.619$ &$10.272$ & $10.369$\\ \hline
\cite{HH-Narison}& $3.48$ & &$9.94$ & \\ \hline
 \cite{HH-Zhang}& $4.26$ &$4.25$ &$9.78$ & $9.85$\\ \hline
  \cite{PDG}& $3.5189$ &? &? & ?\\ \hline
 This work&   $3.57\pm0.14$  &$3.71\pm0.14$ &$10.17\pm0.14$ & $10.32\pm0.14$\\ \hline
 \cite{HH-Roberts}$^*$& $3.910$ &$4.046$ &$10.493$ & $10.616$ \\ \hline\hline
\end{tabular}
\end{center}
\caption{ The masses $M(\rm{GeV})$   of the doubly heavy baryon
states, where the star $*$ denotes the masses of the negative parity
doubly heavy baryon states predicted by the non-relativistic quark
model.}
\end{table}

\begin{table}
\begin{center}
\begin{tabular}{|c|c|c|c|c|}
\hline\hline  & $\Xi_{cc}$& $\Omega_{cc}$&
$\Xi_{bb}$&$\Omega_{bb}$\\\hline
   $\lambda_{+}\,[\rm{GeV}^3]$&   $0.115\pm0.027$  &$0.138\pm0.030$ &$0.252\pm0.064$ & $0.311\pm0.077$\\ \hline\hline
\end{tabular}
\end{center}
\caption{ The  pole residues $\lambda_{+}$ of the doubly
 heavy baryon states.}
\end{table}

\begin{figure}
 \centering
 \includegraphics[totalheight=5cm,width=6cm]{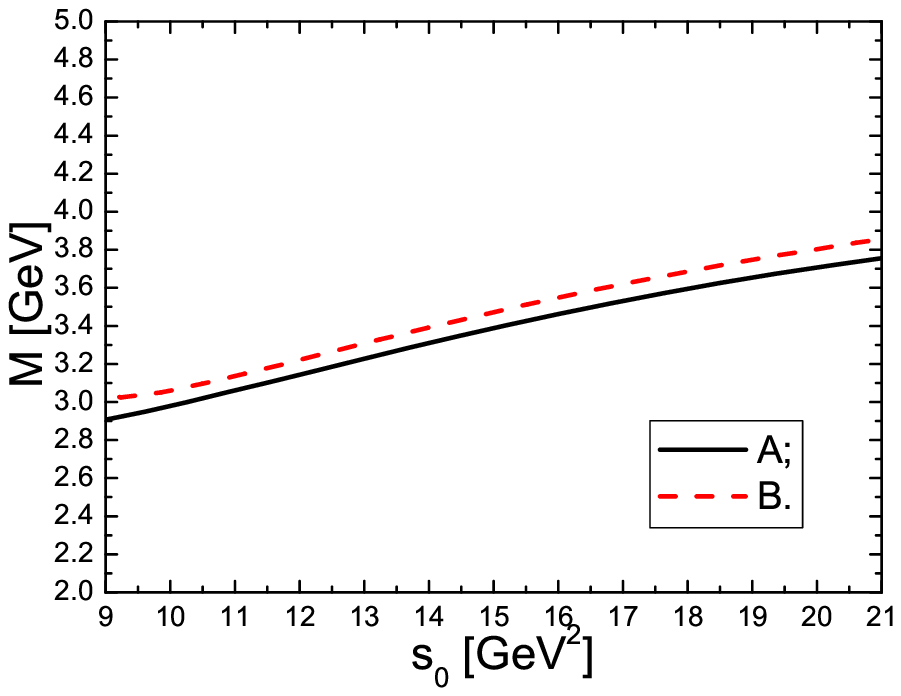}
\includegraphics[totalheight=5cm,width=6cm]{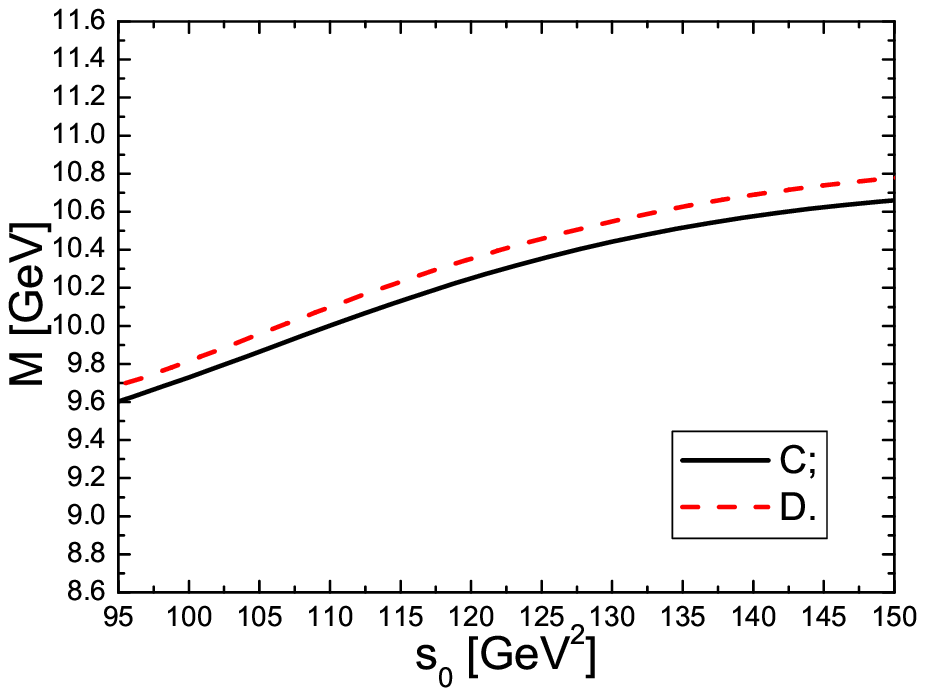}
        \caption{ The masses  $M$ of the doubly  heavy baryon states with variations  of the threshold parameters  $s_0$,
 the $A$, $B$, $C$ and $D$  correspond
      to the channels $\Xi_{cc}$, $\Omega_{cc}$, $\Xi_{bb}$ and $\Omega_{bb}$ respectively, the Borel parameters $T^2$ are taken
to be the central values. }
\end{figure}

\begin{figure}
 \centering
 \includegraphics[totalheight=5cm,width=6cm]{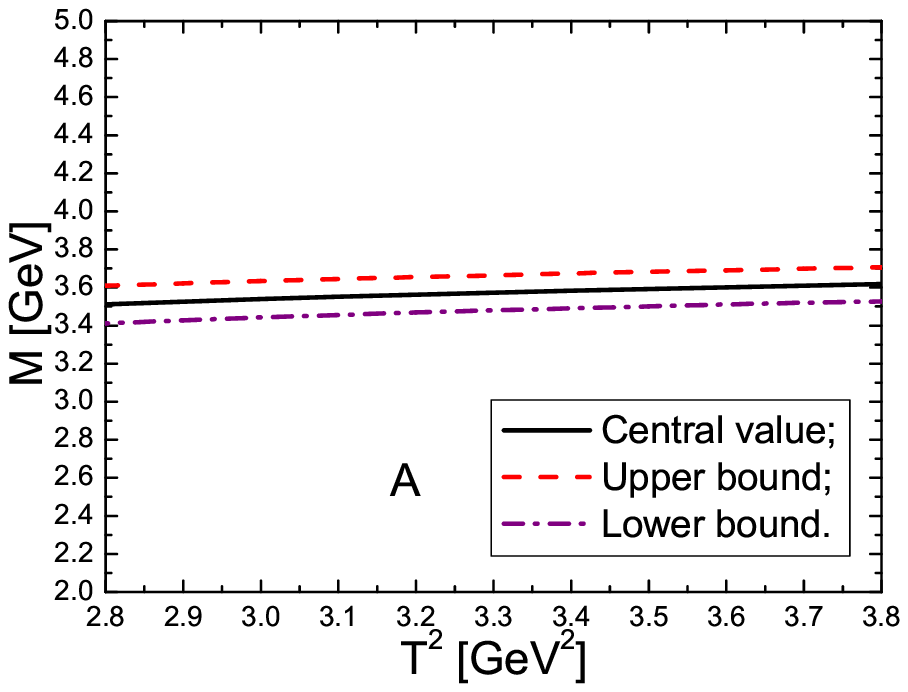}
 \includegraphics[totalheight=5cm,width=6cm]{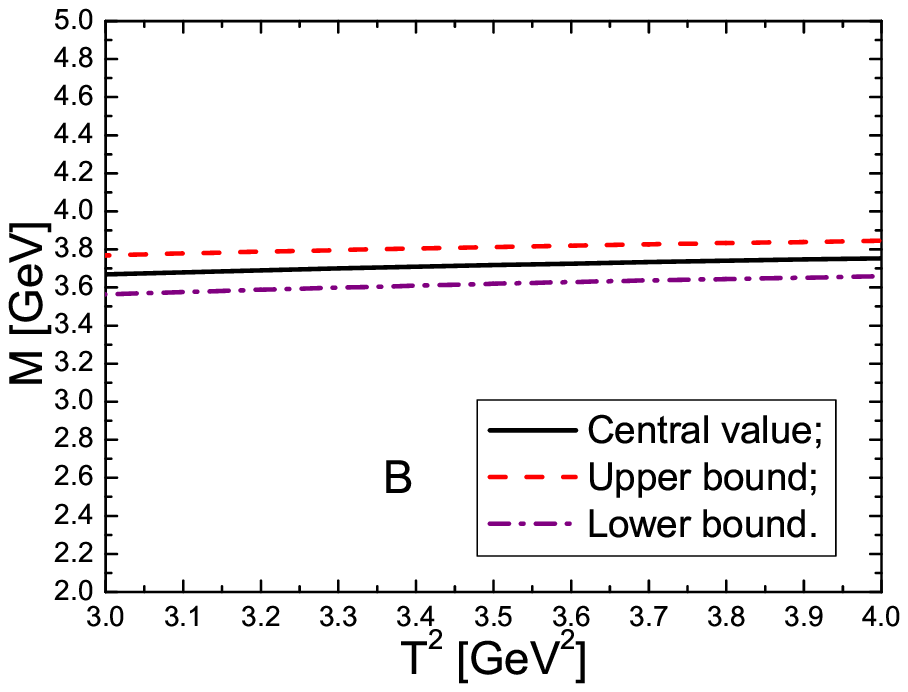}
\includegraphics[totalheight=5cm,width=6cm]{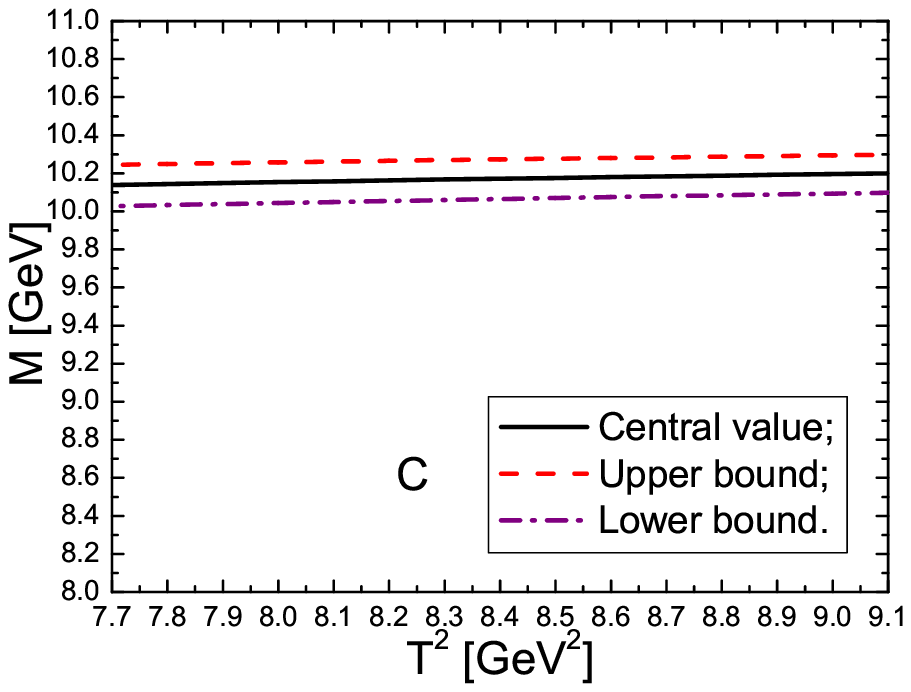}
\includegraphics[totalheight=5cm,width=6cm]{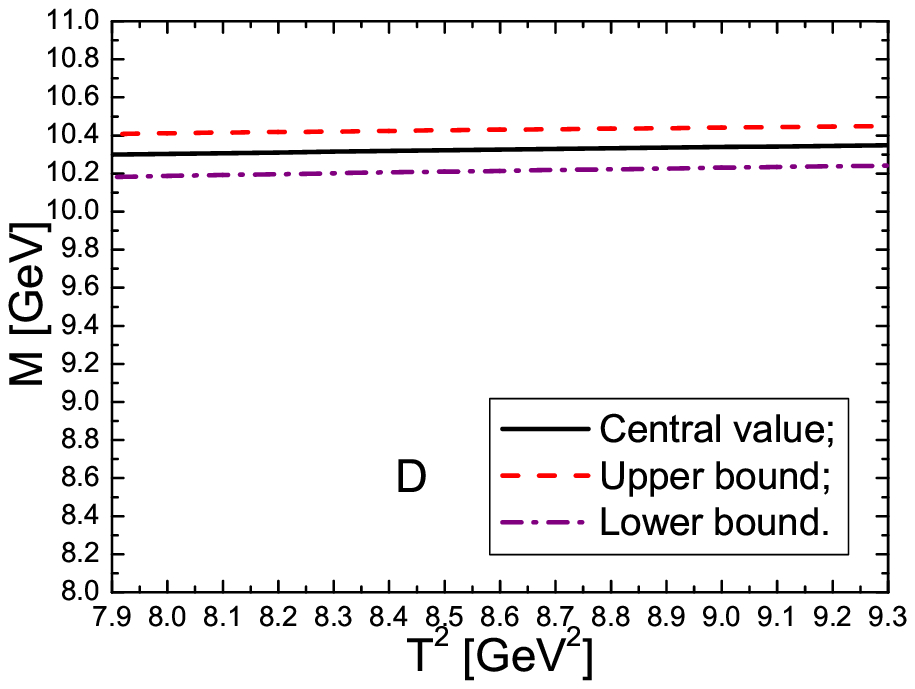}
        \caption{ The masses  $M$ of the double heavy baryon states, the $A$, $B$, $C$ and $D$  correspond
      to the channels $\Xi_{cc}$, $\Omega_{cc}$, $\Xi_{bb}$ and $\Omega_{bb}$ respectively.  }
\end{figure}

\begin{figure}
 \centering
 \includegraphics[totalheight=5cm,width=6cm]{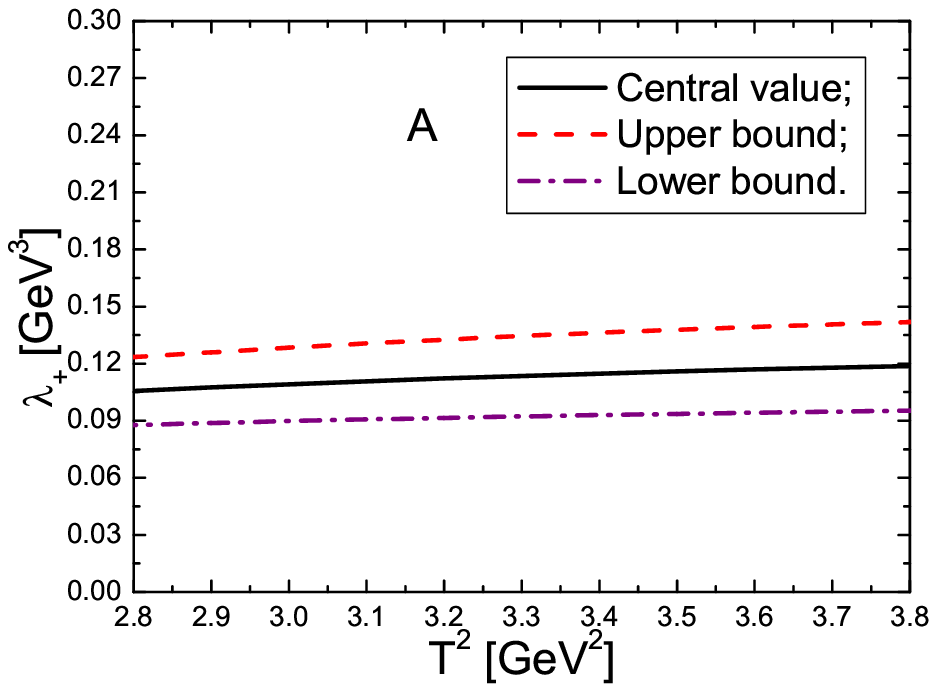}
 \includegraphics[totalheight=5cm,width=6cm]{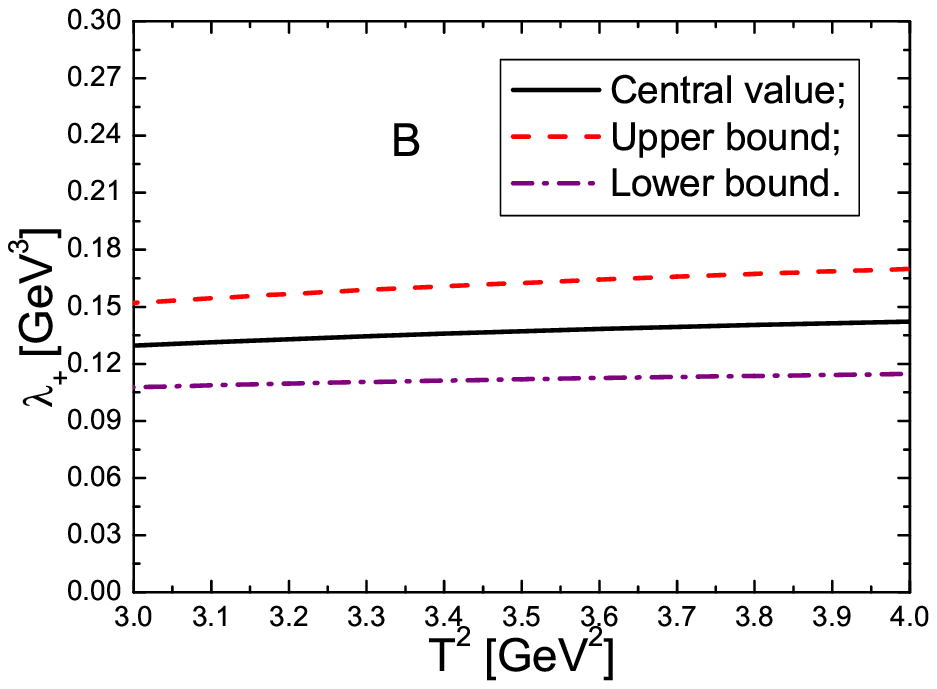}
\includegraphics[totalheight=5cm,width=6cm]{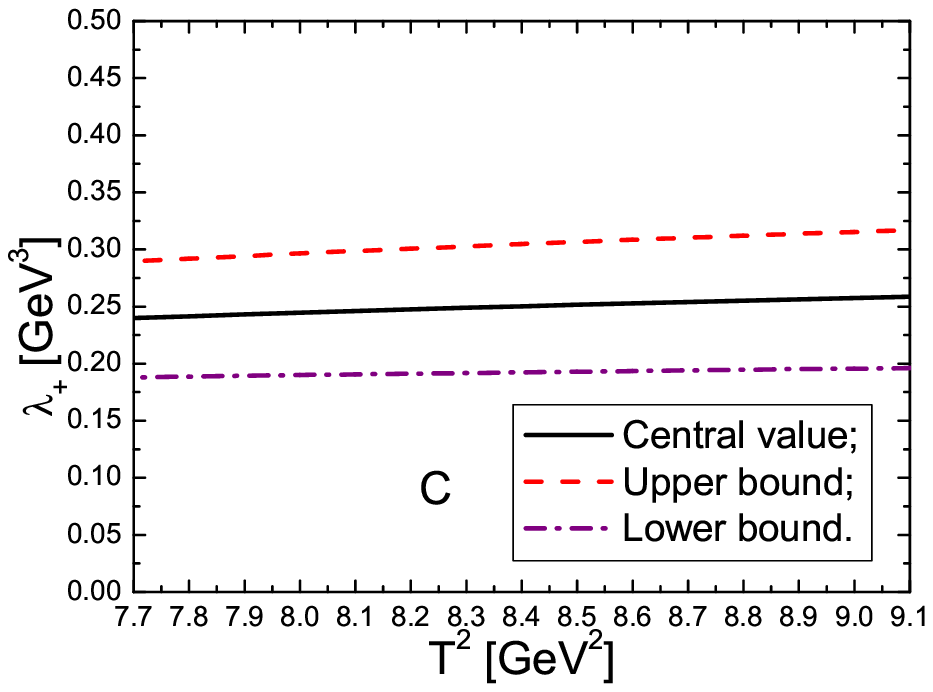}
\includegraphics[totalheight=5cm,width=6cm]{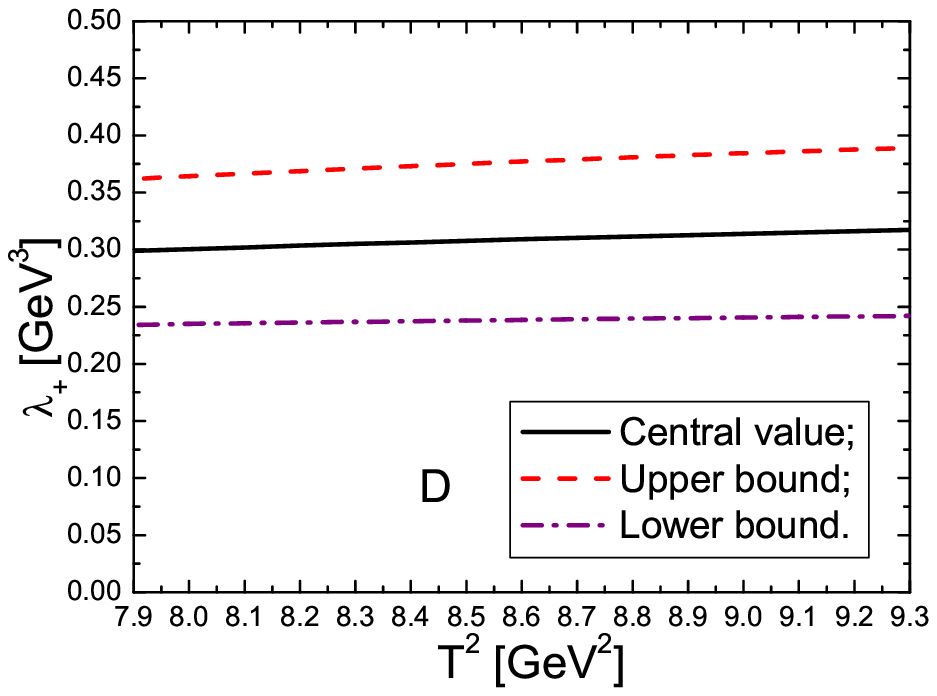}
        \caption{ The pole residues  $\lambda_{+}$ of the doubly heavy baryon states, the $A$, $B$, $C$ and $D$  correspond
      to the channels $\Xi_{cc}$, $\Omega_{cc}$, $\Xi_{bb}$ and $\Omega_{bb}$ respectively.  }
\end{figure}

\begin{figure}
 \centering
 \includegraphics[totalheight=5cm,width=6cm]{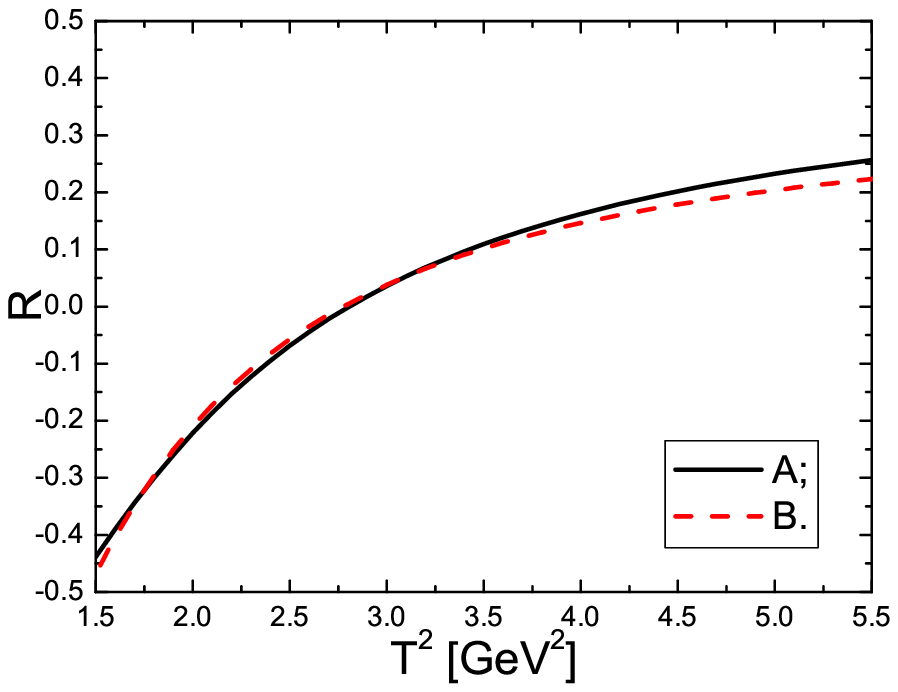}
\includegraphics[totalheight=5cm,width=6cm]{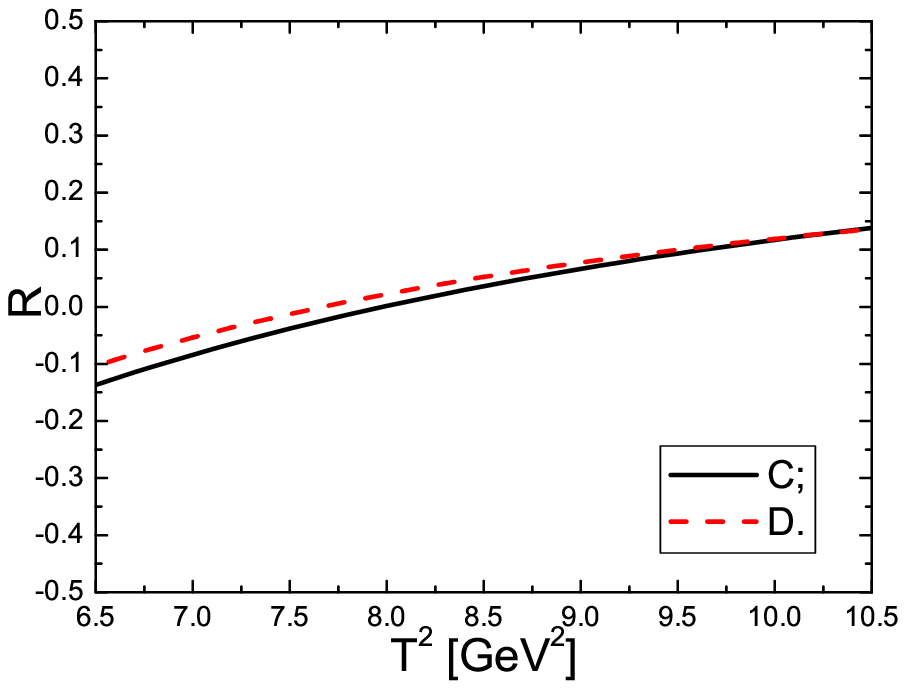}
        \caption{ The ratios between the contributions from the negative parity and positive parity
 doubly  heavy baryon states with variations  of the Borel parameters  $T^2$,
 the $A$, $B$, $C$ and $D$  correspond
      to the channels $\Xi_{cc}$, $\Omega_{cc}$, $\Xi_{bb}$ and $\Omega_{bb}$ respectively. }
\end{figure}

The fractions
\begin{eqnarray}
 R&=& \frac{\int_{\Delta}^{\sqrt{s_0}}dp_0\left[\rho^A(p_0) -\rho^B(p_0)\right]\exp\left[-\frac{p_0^2}{T^2}\right]}
{\int_{\Delta}^{\sqrt{s_0}}dp_0\left[\rho^A(p_0)
+\rho^B(p_0)\right]\exp\left[-\frac{p_0^2}{T^2}\right]}
\end{eqnarray}
 are shown explicitly in Fig.5.  At the value
$T^2=(2.5-4.0)\,\rm{GeV}^2$, $R=(-7\sim17)\%$ in the doubly charmed
baryon channels; and at the value $T^2=(7.5-9.5)\,\rm{GeV}^2$,
$R=(-4\sim10)\%$ in the doubly bottom baryon channels. The
contributions from the negative parity doubly heavy baryon states
are not very large, although they are considerable in some sense. In
this article, the central values of the threshold parameters are
$s_0=4.2\,\rm{GeV}$, $4.3\,\rm{GeV}$, $10.8\,\rm{GeV}$ and
$10.9\,\rm{GeV}$ in  the channels $\Xi_{cc}$, $\Omega_{cc}$,
$\Xi_{bb}$ and $\Omega_{bb}$ respectively, which are larger than the
 masses of the corresponding negative parity doubly heavy baryon
states, see Table 2. If we take the tensor structures $\!\not\!{p}$
and $1$, the contributions from the negative parity baryon states
are included in. In the case of the doubly bottom channels, the
threshold parameters are slightly larger than the thresholds of  the
corresponding  negative parity baryon states, the contaminations are
small, see Fig.5. In fact, without separating the contributions of
the positive parity baryon states from the negative parity baryon
states explicitly, the two criteria (pole dominance and convergence
of the operator product expansion) for choosing  the Borel parameter
$T^2$ and threshold parameter $s_0$ in the conventional QCD sum
rules do not work efficiently; we maybe (or maybe not)  choose the
Borel windows where the contaminations from the negative parity
doubly  heavy baryon states are large.
 With suitable Borel parameters, we can
choose the tensor structures $\!\not\!{p}$ or $1$
  to study the masses and pole residues freely. If we choose the
tensor structure $\gamma_0+1$, the contaminations from the negative
parity doubly heavy baryon states are excluded explicitly.

Those doubly heavy baryon states $\Xi_{cc}$, $\Omega_{cc}$,
$\Xi_{bb}$ and $\Omega_{bb}$ maybe observed at the Tevatron, the
LHCb and the PANDA, especially at the LHCb. For example, the
$\Xi_{cc}$ and $\Omega_{cc}$ can be produced at the high energy $pp$
or $p\bar{p}$ collisions through  the gluon-gluon fusion mechanism
and the intrinsic charm mechanisms, $g+g\to(cc)[^3S_1]_{\bar
3}+\bar{c}+\bar{c}$, $g+g\to(cc)[^1S_0]_6+\bar{c}+\bar{c}$, $g+c\to
(cc)[^3S_1]_{\bar 3}+\bar{c}$, $g+c\to (cc)[^1S_0]_6+\bar{c}$,
$c+c\to (cc)[^3S_1]_{\bar 3}+g$, $c+c\to (cc)[^1S_0]_6+g$,   where
the $(cc)[^3S_1]_{\bar 3}$ (in color anti-triplet {\bf $\bar 3$})
and $(cc)[^1S_0]_6$ (in color sextet $\bf{6}$) are two possible
$S$-wave configurations of the doubly charmed diquark pair $(cc)$
inside the baryon states $\Xi_{cc}$ and $\Omega_{cc}$
\cite{Chang-1,Chang-2,Comput,Baranov96,Berezhnoy96}. The LHCb is a
dedicated $b$ and $c$-physics precision experiment at the LHC (large
hadron collider). The LHC will be the world's most copious source of
the $b$ hadrons, and  a complete spectrum of the $b$ hadrons will be
available through gluon fusion. Furthermore, once reasonable values
of the pole residues $\lambda_{\Omega}$ and $\lambda_{\Xi}$ are
obtained, we can take them as   basic input parameters and study the
revelent hadronic processes  with the QCD sum rules.

\section{Conclusion}
In this article, we study the  ${1\over 2}^+$ doubly heavy baryon
states $\Omega_{QQ}$ and $\Xi_{QQ}$   by subtracting the
contributions from the corresponding ${1\over 2}^-$ doubly heavy
baryon states with the QCD sum rules, and make reasonable
predictions for their masses.  Those doubly heavy baryon states
maybe observed at the Tevatron, the LHCb and the PANDA, especially
at the LHCb. Once reasonable values of the pole residues
$\lambda_{\Omega}$ and $\lambda_{\Xi}$ are obtained, we can take
them as   basic input parameters and study the revelent hadronic
processes with the QCD sum rules.

\section*{Acknowledgements}
This  work is supported by National Natural Science Foundation,
Grant Number 10775051, and Program for New Century Excellent Talents
in University, Grant Number NCET-07-0282, and the Fundamental
Research Funds for the Central Universities.

\end{document}